\documentclass{elsart}

\usepackage{natbib}

 \usepackage{epsfig}

\usepackage{amssymb}

\begin{document}

\begin{frontmatter}



\title{Galactic Structure and Radioactivity Source Distributions}


\author{Ronald Drimmel}

\address{Turin Astronomical Observatory}

\begin{abstract}
A probable sky map of the emission from a short-lived isotrope
produced by massive stars is presented. The model
is based on the nonaxisymmetric component of a dust distribution model
developed to reproduce Galactic FIR emission. Features seen in COMPTEL
observations are qualitatively reproduced.
\end{abstract}


\end{frontmatter}

\section{Introduction}


The opening of the gamma ray window presents us with the possibility 
of tracing star formation (SF) on a Galactic scale via short-lived isotopes
produced by supernovae (SN).  An example is the isotope $^{26}$Al, whose
half-life is short (0.75Myr) and is believed to be primarily produced
by Type II SN, with progenitors of mass $> 10M_\odot$, and WR stars of
mass $> 20M_\odot$. 
The correlation between the distribution of $^{26}$Al 
and SF activity derives from 
the brief lifetimes of these massive stars, typically
shorter than 20Myr, and the short half-life of $^{26}$Al, 
the latter insuring that the emitting material does
not have time to turbulently mix throughout the ISM. 
Empirical support is found in the 
observed correlation between $^{26}$Al emission and millimeter radiation
\citep{Knod99a}, while $^{26}$Al emission has been used to place
theoretical constraints on the global SF rate \citep{Knod99aa, TDH97}.

To date the only gamma ray all-sky survey available is that
from the {\em Compten Gamma-Ray Observatory}, which produced a low
resolution sky map of MeV radiation showing irregular emission along
the Galactic plane. In the future these observations will be
supplemented by the INTEGRAL mission. 
The purpose of this paper is to give a description of the anticipated 
Galactic emission from short-lived isotopes, as seen from the Sun's
position in the Milky Way, and as inferred from a model of the Galactic
distribution of dust and stars. 

\section{Star formation on a Galactic scale}

Spiral arms can be regarded as star formation fronts. 
While SF does not take place exclusively in spiral arms,
they can be regarded as the principal sites of SF activity
in most disk galaxies.
Indeed, it is for this reason that they are visible at optical wavelengths
as complexes of HII regions. Our position within the Milky Way
impedes a complete mapping of the it's spiral arms, but observations
of Galactic HII regions have allowed a partial tracing of the spiral
arms on our side of the Galaxy \citep{GG76, TC93}


This tracing of the Milky Way's spiral arms is found to 
trace well the far-infrared (FIR) emission associated with spiral arms
\citep{DS01}. The Galactic FIR emission was
observed by the COBE satellite, and is due to interstellar dust,
entrained in the gaseous component of the Galactic disk. The time
scales for Galactic dust production and 
redistribution is considerably longer than the
dynamical time scales that produces the concentration of gas and dust 
into the spiral arms, making the dust a good tracer of the azimuthal variation
of the gas as well. It is therefore no surprise that a tracing of
the spiral arms based on the location of HII regions can be used to
model the FIR emission, since SF preferentially occurs
where the gas density is highest. 

For our purposes here I adopt the nonaxisymmetric component of the
dust density model, used to reproduce the COBE FIR observations, as a
tentative model of the probable flux density from a short-lived
isotope produced by massive stars, such as $^{26}$Al. For details
of the
model, please refer to \citet{DS01}. The nonaxisymmetric
component is shown in Figure 1, showing clearly the spiral arms. 
The reader should also note another feature near the Sun,
a small dust lane that corresponds to the Orion ``arm''. 

\begin{figure*}
\epsfbox{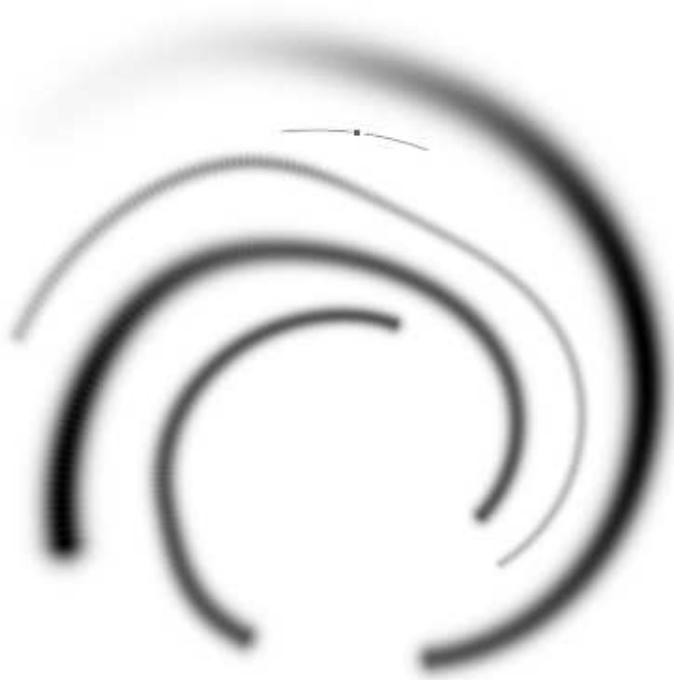}
\caption{Map of the surface density of the 
dust associated with the spiral arms. The Sun's
position is indicated as the block dot, upper center. The map is
incomplete on the side of the Galaxy opposit the Sun due to lack of
HII data.}
\label{f1}
\end{figure*}

I have argued above that the nonaxisymmetric component of the dust
distribution is logically associated with SF, so using it to
describe $^{26}$Al emission is strictly valid only if SF and 
$^{26}$Al production are contemporaneous, as
Galactic differential rotation will shear the distribution of
newly born stars with respect to the spiral arm pattern.
However,  as the lifetimes of $^{26}$Al producers are much
shorter than the period of Galactic rotation, the resulting azimuthal offsets
are negligible over most of the Galaxy, as shown in Figure 2.

\begin{figure*}
\epsfysize=12cm
\epsfbox{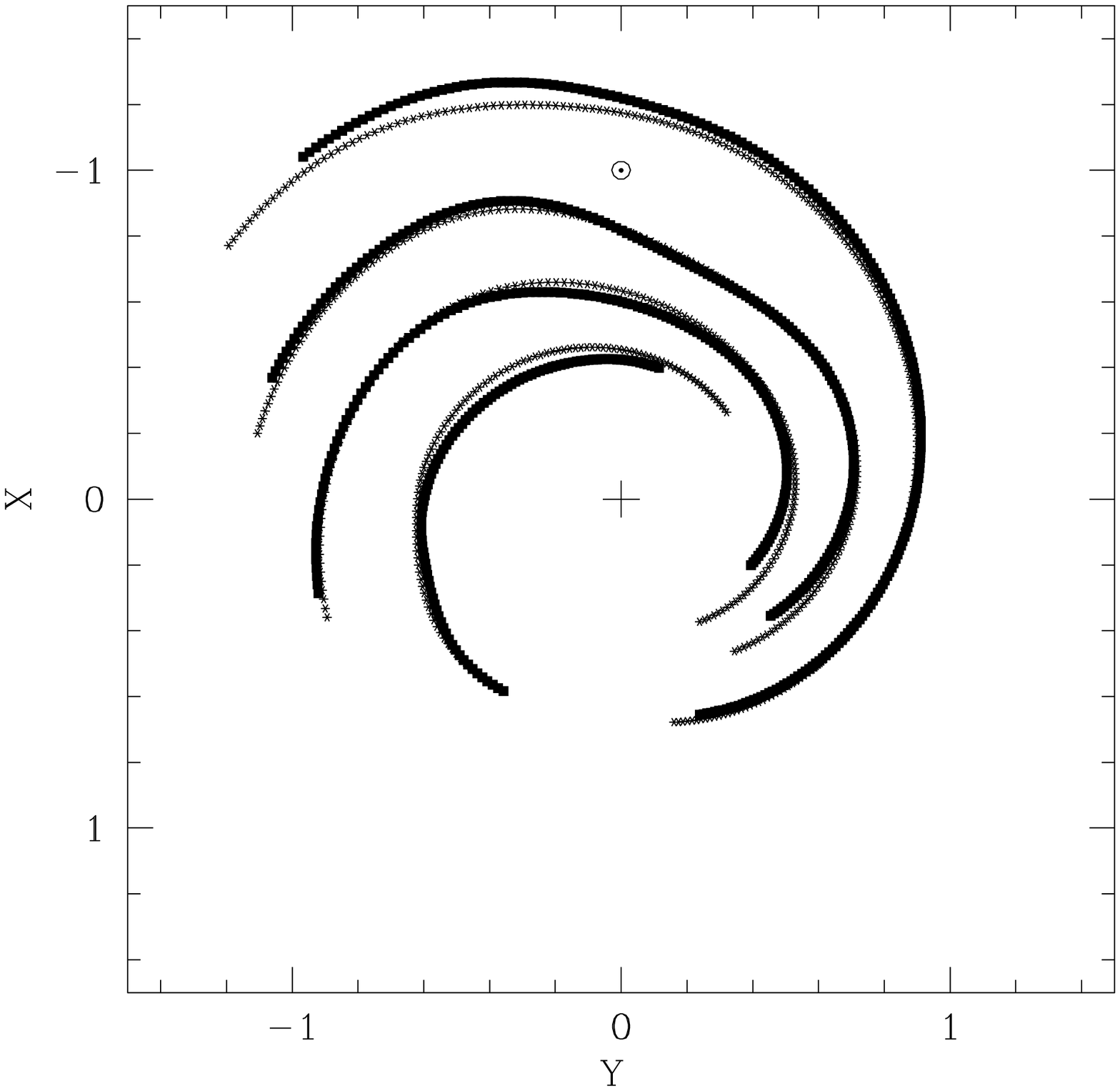}
\caption{Schematic diagram showing the geometry of the spirals arms as
currently seen in the dust (thick solid line) and the geometry resulting after
20Myr of shearing by differential Galactic rotation (astericks),
assuming $R_\odot=8$ kpc, corotation at $0.83R_\odot$, and a flat rotation
curve of 200 km/s.}
\label{f2}
\end{figure*}

While we can have some confidence that azimuthal variation of the dust
should correspond to that of $^{26}$Al emission, we are left with the
question of the radial variation. A priori, it is not obvious that
the radial variation in the dust and $^{26}$Al will be the same, one being
the product of past SF and dynamical evolution, while the other is
a result of current SF. Nevertheless it is worth mentioning that the
adopted model for the radial variation of the dust density in the
spiral arms is of the same form as the free electron density associated
with the arms \citep{TC93}. As this electron density is a
consequence of the ionizing radiation of young OB stars, it should
also reflect the radial variation of SF activity.

\section{Expected sky emission from short-lived isotopes}

Having a presumed three-dimensional model of the $^{26}$Al at our
disposal, it is 
now possible to produce a sky map of the expected emission from this
isotope, the flux density being proportional to the density of the
emitting material. Another approximation made is similar to that made
for the FIR, that at the wavelengths (energies) in question, the
Galaxy is optically thin. Figure 3 shows the resulting sky map.
Several features are worth pointing out in this figure. First, within
90 degrees of the Galactic center the spiral arms are seen as emission
peaks along the Galactic plane. Particularly visible are the tangents at
Galactic longitudes $-$75 and $-$50 degrees, corresponding to the
Sagittarius-Carina and Scutum arms respectively. At Galactic longitudes
of approximately 90. and $-$100. two more bright spots are notable for
their extent in Galactic latitudes; these are due to the local Orion
``arm'', a structure important in our sky only because of our
proximity to it. Lastly, emission  
between $\pm 90$ degrees is attributed to the Perseus
arm, whose scale height is significantly larger than the other spiral
arms. Another nonaxisymmetric structure that is less obvious in the
sky map is the Galactic warp, which causes the Perseus arm to deviate
from the Galactic plane.

\begin{figure*}
\epsfxsize=17cm
\epsfbox{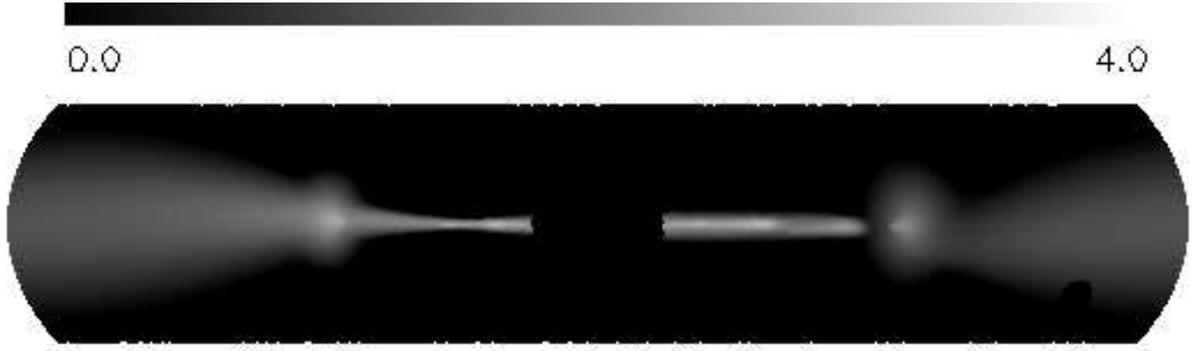}
\caption{Sky map of the emission resulting from the spiral arms shown
in Figure 1. Emission within 20 degrees of the Galactic center is not shown.}
\label{f3}
\end{figure*}

To better see the variation in the Galactic plane Figure 4 shows the 
emission profile at $b=0$. The spiral arms are easily seen by the
sawtooth pattern they produce in the profile. This characteristic
profile was also predicted by \citet{Prant93} with a very similar
model, using the same spiral arm geometry.
I also point out that the incompleteness of the spiral arm map on the
other side of the Galaxy is not important for producing the sky maps,
as their small scale height insure that they remain unresolved.

\begin{figure*}
\epsfxsize=15cm
\epsfbox{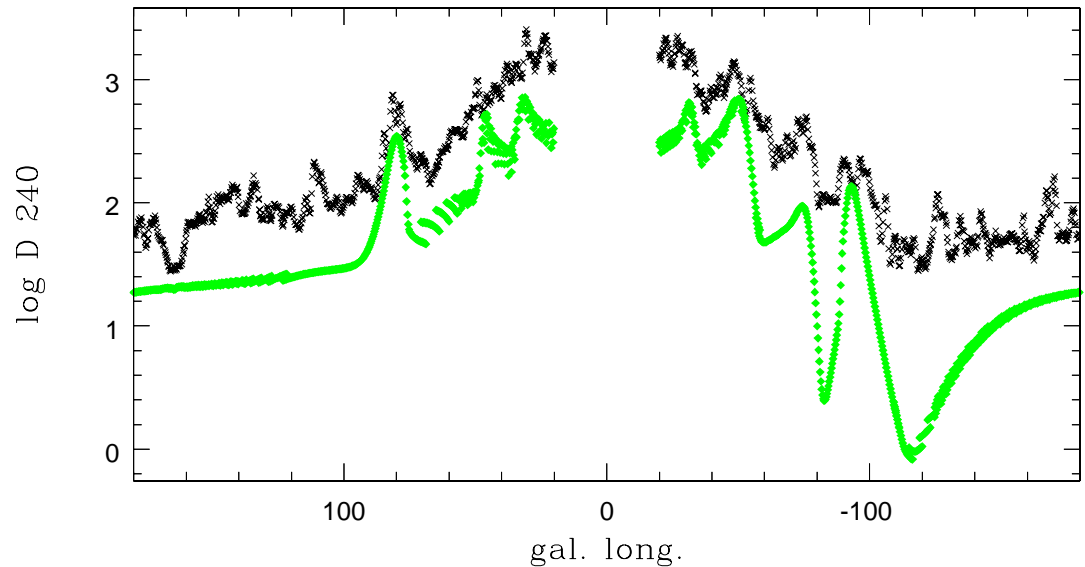}
\caption{Modelled $^{26}$Al emission profile (arbitrary units) in the Galactic
plane (diamonds), plotted with the $240 \mu$ profile (MJy sr$^{-1}$)
observed by COBE for comparison. }
\label{f4}
\end{figure*}

\section{Summary}


Using a model based on FIR observations, a predicted map of the
emission from short-lived isotopes has been presented at resolutions
comparable to the future INTEGRAL mission though, not being a survey
mission, it will not produce such a sky map. This model is
parametric, using smooth mathematical functions to describe the
distribution of the emitting material, and is thus much simpler than
reality. This can be immediately seen by comparing Figure 3
with the COMPTEL sky map at 1.8MeV
\citep[][or see http://cossc.gsfc.nasa.gov/cossc/comptel/]{Knod99b}.
However, the major features 
in the map at Galactic longitudes $> 30$ degrees are reproduced on a
qualitative level. This includes the location of the spiral arm
tangents, the two bright spots attributable to the local Orion
``arm'', and the contribution of the Perseus arm to emission toward
the Galactic anticenter with its relatively large range in Galactic
latitude. 

No attempt has been made here to fit the model to actual
gamma-ray data; for this the reader is referred to \citet{Knod96},
who uses the same spiral arm geometry, adopting the
free-electron density model of \citet{TC93} as a template. Perhaps the
only significant differences between this model and the one presented
here is a reduction factor on the Sagittarius-Carina arm, the
Galactic warp and a flair in the spiral arm scale height.

As a worker in the field of Galactic structure, 
preparing this report underlined for me
the value of gamma ray mapping missions, as it provides an avenue for
observing and identifying the distribution of SF regions on a large
scale. This is important for untangling the contribution from bright
young sources which contaminates other wavelengths and frustrates
efforts at mapping the mass distribution in the Galactic disk.


\end{document}